\documentclass[conference]{IEEEtran}
\IEEEoverridecommandlockouts
\usepackage{url}
\usepackage{cite}
\usepackage{amsmath,amssymb,amsfonts}
\usepackage{algorithmic}
\usepackage{graphicx}
\usepackage{textcomp}
\usepackage{comment}
\usepackage[hidelinks]{hyperref}
\usepackage[dvipsnames]{xcolor}
\usepackage{subcaption}
\usepackage{multirow}
\usepackage{multicol}
\usepackage{amssymb}
\newcommand{\cmark}{\ding{51}}%
\newcommand{\xmark}{\ding{55}}%
\usepackage{lipsum} 
\usepackage{mdframed} 
\usepackage{tabularx}
\usepackage{flushend}
\usepackage{pifont}
\usepackage{listings}
\definecolor{codegreen}{rgb}{0,0.6,0}
\definecolor{codegray}{rgb}{0.5,0.5,0.5}
\definecolor{codepurple}{rgb}{0.58,0,0.82}
\definecolor{backcolour}{rgb}{0.95,0.95,0.92}
\makeatletter
\newcommand\notsotiny{\@setfontsize\notsotiny{7}{7}}
\makeatother

\lstdefinestyle{mystyle}{
    commentstyle=\color{codegreen},
    keywordstyle=\color{magenta},
    numberstyle=\tiny\color{codegray},
    stringstyle=\color{codepurple},
    basicstyle=\ttfamily\notsotiny,
    breakatwhitespace=false,         
    breaklines=true,                 
    captionpos=b,                    
    keepspaces=true,                 
    showspaces=false,                
    showstringspaces=false,
    showtabs=false,                  
    tabsize=1
}

\newcolumntype{?}{!{\vrule width 1.5pt}}

\newcommand{\exampleone}[1]{%
 \par\vspace{1pt}%
  \noindent\textbf{Example 1.} \textit{#1}%
   \par\vspace{1pt}%
}

\newcommand{\exampletwo}[1]{%
 \par\vspace{1pt}%
  \noindent\textbf{Example 2.} \textit{#1}%
   \par\vspace{1pt}%
}

\newcommand{\examplethree}[1]{%
 \par\vspace{1pt}%
  \noindent\textbf{Example 3.} \textit{#1}%
   \par\vspace{1pt}%
}

\usepackage{siunitx}
\def\BibTeX{{\rm B\kern-.05em{\sc i\kern-.025em b}\kern-.08em
    T\kern-.1667em\lower.7ex\hbox{E}\kern-.125emX}}
\begin{document}

\title{Towards an Automatic Framework for Solving Optimization Problems with Quantum Computers
}

\author{\IEEEauthorblockN{Deborah Volpe\IEEEauthorrefmark{1},
Nils Quetschlich\IEEEauthorrefmark{2}, Mariagrazia Graziano\IEEEauthorrefmark{3},  Giovanna Turvani\IEEEauthorrefmark{1}, and Robert Wille\IEEEauthorrefmark{2}\IEEEauthorrefmark{4} }
\IEEEauthorblockA{\IEEEauthorrefmark{1}Department of Electronics and Telecommunications,
Politecnico di Torino
Italy\\
\IEEEauthorrefmark{2}Chair for Design Automation, Technical University of Munich, Germany\\
\IEEEauthorrefmark{3}Department of Applied Science and Technology,
Politecnico di Torino
Italy\\
\IEEEauthorrefmark{4}Software Competence Center Hagenberg GmbH (SCCH), Austria\\
 \href{mailto:deborah.volpe@polito.it}{deborah.volpe@polito.it},
\href{mailto:nils.quetschlich@tum.de}{nils.quetschlich@tum.de},
\href{mailto:mariagrazia.graziano@polito.it}{mariagrazia.graziano@polito.it},\\
\href{mailto:giovanna.turvani@polito.it}{giovanna.turvani@polito.it} and \href{mailto:robert.wille@tum.de}{robert.wille@tum.de}}
\vspace{-32pt}}

\maketitle
\begin{abstract}
Optimizing objective functions stands to benefit significantly from leveraging quantum computers, promising enhanced solution quality across various application domains in the future. However, harnessing the potential of quantum solvers necessitates formulating problems according to the \textit{Quadratic Unconstrained Binary Optimization} (QUBO) model, demanding significant expertise in quantum computation and QUBO formulations. This expertise barrier limits access to quantum solutions.

Fortunately,  automating the conversion of conventional optimization problems into QUBO formulations presents a solution for promoting accessibility to quantum solvers. This article addresses the unmet need for a comprehensive automatic framework to assist users in utilizing quantum solvers for optimization tasks while preserving interfaces that closely resemble conventional optimization practices. The framework prompts users to specify variables, optimization criteria, as well as validity constraints and, afterwards, allows them to choose the desired solver. Subsequently, it automatically transforms the problem description into a format compatible with the chosen solver and provides the resulting solution. Additionally, the framework offers instruments for analyzing solution validity and quality. 

Comparative analysis against existing libraries and tools in the literature highlights the comprehensive nature of the proposed framework. Two use cases (the knapsack problem and linear regression) are considered to show the completeness and efficiency of the framework in real-world applications.

Finally, the proposed framework represents a significant advancement towards automating quantum computing solutions and widening access to quantum optimization for a broader range of users. 

The framework is publicly available on GitHub (\url{https://github.com/cda-tum/mqt-qao}) as part of the Munich Quantum Toolkit (MQT).

\end{abstract}

\vspace{-3pt}
\section{Introduction}
\vspace{-3pt}
\textit{Quantum computers} have the potential to enhance the optimization of \textit{objective functions} in specific application domains such as machine learning~\cite{date2021qubo, hu2019quantum}, scheduling~\cite{zhang2022solving}, and resource allocation~\cite{ohyama2023resource}. However, leveraging \textit{quantum solvers} requires writing the problem in a suitable format, demanding proficiency in quantum computers and \mbox{quantum-compliant} problem formulations. This challenges researchers or industries not directly engaged in quantum computation, inhibiting the initial exploration of quantum solutions for \mbox{real-world} use cases, such as linear regression~\cite{date2021qubo}. Automating the transformation of problems into a \mbox{quantum-compliant} format offers a solution to this obstacle.

This article proposes a framework to assist \textit{\mbox{non-experts}} in exploring the \textit{potential of quantum solvers} for \textit{optimization problems}. It keeps the interfaces for the problem specification as similar as possible to conventional optimization. Users are prompted to declare variables and their operational ranges, optimization criteria, as well as validity constraints, and afterwards, allows them to choose the desired solver. Subsequently, the tool automatically transforms the problem description into a compatible format for the selected solver and delivers the obtained solution. Moreover, it offers mechanisms for evaluating solution validity and quality.

The framework's characteristics are compared to those of libraries and analogous tools already in the literature, distinguishing itself for its \textit{comprehensiveness}. To this end, two use cases (knapsack problems and linear regression models) are considered to demonstrate the framework's effectiveness and flexibility in real-world application contexts. Notable benefits are observed from the user's perspective in both cases. Consequently, this framework, which is publicly available on GitHub (\url{https://github.com/cda-tum/mqt-qao}) as part of the Munich Quantum Toolkit (MQT) \cite{willeMQTHandbookSummary2024}, represents an important step toward automating quantum computing solutions.

The rest of the article is organized as follows. 
Section~\ref{sec:QuantumOptimization} briefly reviews quantum optimization, focusing on \mbox{quantum-compliant} problem formulations, and discusses the current workflow for solving an optimization problem with a quantum solver. 
Based on that, Section~\ref{sec:Towards} outlines the automation opportunities, which constitute the motivation behind the framework, and a comparison with the \mbox{state-of-the-art}. The actual implementation and its characteristics are described in Section~\ref{sec:Implementation}. The use cases are presented in Section~\ref{sec:usecases}, evaluating the framework from a user perspective and discussing its effectiveness and results. Finally, in Section~\ref{sec:conclusions}, conclusions are drawn, and future perspectives are illustrated.
\vspace{-3pt}
\section{Solving Optimization Problems\\ with Quantum Computers}\label{sec:QuantumOptimization}
\vspace{-3pt}
This section provides a comprehensive overview of quantum optimization, focusing on solvers and the QUBO formulation, which is recognized as the best model for exploiting quantum computing. Additionally, the key steps required for leveraging quantum computers to solve optimization problems are outlined. In particular, stages demanding quantum and QUBO expertise are emphasized since the proposed framework aims to automate these.
\vspace{-5pt}
\subsection{Overview} \label{sec:Overview}
\vspace{-3pt}
In order to exploit the quantum computing paradigm for optimization problems, there are two possibilities: employing a \textit{Quantum Annealer} (QA)---a \textit{\mbox{special-purpose} quantum computer} theorized in 1998~\cite{kadowaki1998quantum,chakrabarti2023quantum,rajak2023quantum, venturelli2015quantum, ManufacturedSpins}, using the natural properties of a quantum system evolution for obtaining the ground state---or executing algorithms entirely or partially on a \textit{\mbox{general-purpose} quantum computer compliant with the quantum circuit model}~\cite{nielsen_quantum_2010}.

In the second case, nowadays, the quantum computer is usually exploited for accelerating specific tasks also involving classical computers. The most popular approaches in this context are:
 \vspace{-1pt}
\begin{itemize}
    \item \textit{Quantum Approximate Optimization Algorithm} (QAOA, ~\cite{blekos2023review,farhi2014quantum}), which is a hybrid \mbox{quantum-classical} algorithm that approximates the adiabatic evolution of a quantum system encoding the optimization problem in the Hilbert space; 
    \item \textit{Variational Quantum Eigensolver} (VQE, ~\cite{tilly2022variational, peruzzo2014variational}), which is a hybrid \mbox{quantum-classical} algorithm for identifying the lower energy eigenstate of a given physical system, allowing its applicability in a combinatorial optimization context;
    \item \textit{Grover Adaptive Search} (GAS, ~\cite{bulger2003implementing, gilliam2021grover, sano2023accelerating, sano2023qubit, giuffrida2022engineering}), a hybrid \mbox{quantum-classical} algorithm, which implements a successive approximation method for estimating the optimal value. It executes the Grover Search algorithm for sampling negative values of the problem cost function, iteratively moving up the classically obtained sample until the last negative value is found.
\end{itemize}
 \vspace{-1pt}
 
The formulation commonly employed for solving optimization problems with quantum computers is the \textit{QUBO} model~\cite{glover2018tutorial,combarro2023practical}. The acronym stands for \textit{Quadratic Unconstrained Binary Optimization}, where \textit{quadratic} indicates the highest degree of the objective function, \textit{unconstrainted} means that the constraints cannot be considered conventionally, \textit{binary} signifies that only unipolar binary variables can be considered, and \textit{optimization} emphasizes the purpose of the model. The associated problem objective function can be written as
\vspace{-8pt}
\begin{equation}
	f(\textbf{b}) = \gamma + \sum_{i}  \alpha_{i}\cdot b_{i}  + \sum_{i<j} \beta_{ij} \cdot b_{i}b_{j} \, ,
  \vspace{-6pt}
\end{equation}
where $b_{i} \in [0,1]$ is a binary variable, $b_{i}b_{j}$ is a coupler that allows two variables to influence each other, $\alpha_{i}$ is a weight or bias associated with a single variable, $\beta_{ij}$ is a strength that controls the influence of variables $i$ and $j$, and  $\gamma$ is an offset that can be neglected during optimization. Usually, this model is employed considering the \textit{minimization} as the optimization direction. \\
This formulation can consider a constraint through the \textit{aggregation method}. In particular, a \textit{quadratic penalty function} $g(x)$---assuming value 0 for configurations satisfying the constraints and a positive amount otherwise---is included in the objective function, as shown by
\vspace{-4pt}
\begin{equation}
\textrm{minimize}  \quad y = f(\textbf{b}) + \lambda g(\textbf{b}) \, ,
	\label{eq:QuboFormulation}
 \vspace{-2pt}
\end{equation}
where $\lambda$ is a positive penalty weight assigned to the constraint quadratic function $g(\textbf{b})$.

Note that the QUBO formulation is also compatible with some classical solvers like \textit{Simulated Annealing} (SA~\cite{Kirkpatrick1983}). This algorithm is commonly employed for the initial validation of a QUBO formulation and as an alternative when the problem size exceeds the capabilities of current \textit{Noisy Intermediate-Scale Quantum} (NISQ) devices and simulators since it can handle thousands of binary variables on classical machines. 

Remembering these concepts, the rest of the section covers the steps required for solving an optimization problem with quantum computers. 


\vspace{-4pt}
\subsection{Problem Specification} \label{sec:ProblemSpec}
\vspace{-3pt}
The first step for leveraging the optimization procedure is the declaration of the \textit{problem specifications}, which consist of the degrees of freedom (i.e., the variables),  the criteria for optimization (described as objective functions), and any requirements (also called constraints) that a valid solution must meet. The problem \textit{variables} can be \textit{binary} (unipolar or bipolar), \textit{discrete} (specifying the valid values), and \textit{continuous} (defining the operative range). Meanwhile, the objective function is a parametric description of a figure of merit whose optimal value could be its minimum or maximum, requiring the specification of the \textit{optimization direction}. In the case of a problem composed of more than one expression to optimize (multi-objective optimization), a \textit{criterion} is essential to establish the relative importance of each contribution in relation to the others. 

These operations do not depend on the problem and the solver, thus maintaining consistency whether moving from classical to quantum optimization context. In the following, two examples, considered representative case studies throughout the remainder of this article, are introduced to illustrate this. The former is characterized by inequality constraints to handle, while the latter involves real variables. Together, they provide a wide overview of possible real-case scenarios. 
\vspace{1pt}
\exampleone{The knapsack problem is a practical challenge where the objective is to determine the optimal subset $S$ from a collection of $N_{\textrm{obj}}$ objects to put into a bag. Each object is characterized by its weight $w_{\textrm{arr}_i}$ and preference score $p_{\textrm{arr}_i}$. The optimal subset maximizes the total score, i.e., the sum of the scores of the taken object $\sum_{\textrm{obj}_i \in S} p_{\textrm{arr}_i}$, and ensures that the total weight $\sum_{\textrm{obj}_i \in S} w_{\textrm{arr}_i}$ does not exceed the maximum~$W_\textrm{max}$. \\
The problem variables represent the objects themselves, and the optimization goal is to maximize the total score. Additionally, a valid solution must adhere to the constraint that the total weight does not exceed $W_\textrm{max}$. 
}
\vspace{1pt}
\exampletwo{Linear regression seeks to determine the straight line that best fits a given dataset by minimizing the Euclidean error function:
\begin{equation*}
  E(w) = \Vert X w + Y \Vert = w^{T}X^{T}X^{T}Xw - 2w^TX^TY+Y^TY\, ,
\end{equation*}
where $X \in \mathbb{R}^{N\times(d+1)}$ is the augmented regression data matrix, $N$ the number of data points in the training set, $d$ the number of features, $Y \in \mathbb{R}^{N}$ is the regression label of the training data, $w \in \mathbb{R}^{d+1}$ the regression weights and $E(w)$ the Euclidian error function.\\
The problem variables are the regression weights $w \in \mathbb{R}^{d+1}$, and the optimization objective is to minimize the Euclidean error function.  This problem is unconstrained. 
}
\vspace{-3pt}
\subsection{Solver Selection}
\vspace{-3pt}
After the problem specification stage, the solving approach has to be chosen among QA, QAOA, VQE, and GAS.\\
Subsequently, accurate consideration must be given to selecting appropriate settings for the chosen solver. For instance, choosing an appropriate annealing time for the problem of interest is crucial when QA is selected.
\vspace{-3pt}
\subsection{Encoding the Problem} \label{sec:writing}
\vspace{-3pt}
After the solver selection, the problem must be written in a compliant format. As previously discussed, all the mentioned approaches support the QUBO formulation, which involves exclusively \textit{binary variables}. Therefore, continuous and discrete variables must be expressed as a weighted set of binary ones through proper \textit{encoding mechanisms}. Additionally, it requires including the constraints in the objective function through \textit{weighted penalties} for allowing their evaluation during the optimization procedure, as explained in Section \ref{sec:Overview}.

Moreover, the quantum solver can only handle the minimization problem, requiring a sign change in the case of problems with a maximization objective, and solve \textit{second-order polynomials}, necessitating a polynomial reduction step.


In the case of \textit{\mbox{multi-objective}} optimization, the \textit{aggregation} approach can be employed by combining
objective functions into a higher scalar one, expressing a preference criterion.

Again, these steps are illustrated by considering the two running examples. 

\vspace{1pt}
\exampleone{When formulating a knapsack problem in QUBO form, assigning a binary variable to each object is necessary. This step is straightforward in this case, as the problem naturally lends itself to a binary representation of variables. Additionally, optimization involves expressing the objective as a unified cost function, incorporating constraints as penalty functions. Translating inequality constraints into penalty functions requires considerable expertise in QUBO formulation, especially due to the need for auxiliary variables.
}
\vspace{1pt}
\exampletwo{The initial step for writing a linear regression problem into a QUBO form involves encoding the variables into a binary format. This process requires a suitable encoding mechanism since the original variables are \mbox{real-valued}. Subsequently, the expression must be redefined as a function of these binary variables.
}
\vspace{-3pt}
\subsection{Solving the Problem}\label{sec:execute}
\vspace{-3pt}
The resulting formulation is then submitted to the chosen solver with a proper \textit{parameters configuration}. Due to their stochastic nature, these solvers are usually run multiple times, and the best-obtained result is considered.

Access to real quantum devices is enabled through cloud services using a dedicated account. Alternatively, QAOA, VQE, and GAS can also be executed on classical simulators, such as those referenced in~\cite{fingerhuth2018open, guerreschi2020intel,zulehner2018advanced, hillmich2020just, vincent2022jet, villalonga2019flexible }.

\vspace{-3pt}
\subsection{Solution Analysis}\label{sec:decode}
\vspace{-3pt}
Finally, the obtained solution must be \textit{decoded}, the original problem variables recovered, and its quality evaluated. This process consists of evaluating the initial cost functions with the found configuration and verifying the satisfaction of constraints. 


\begin{figure*}[t]
	\centering
	\includegraphics[width=0.87\textwidth]{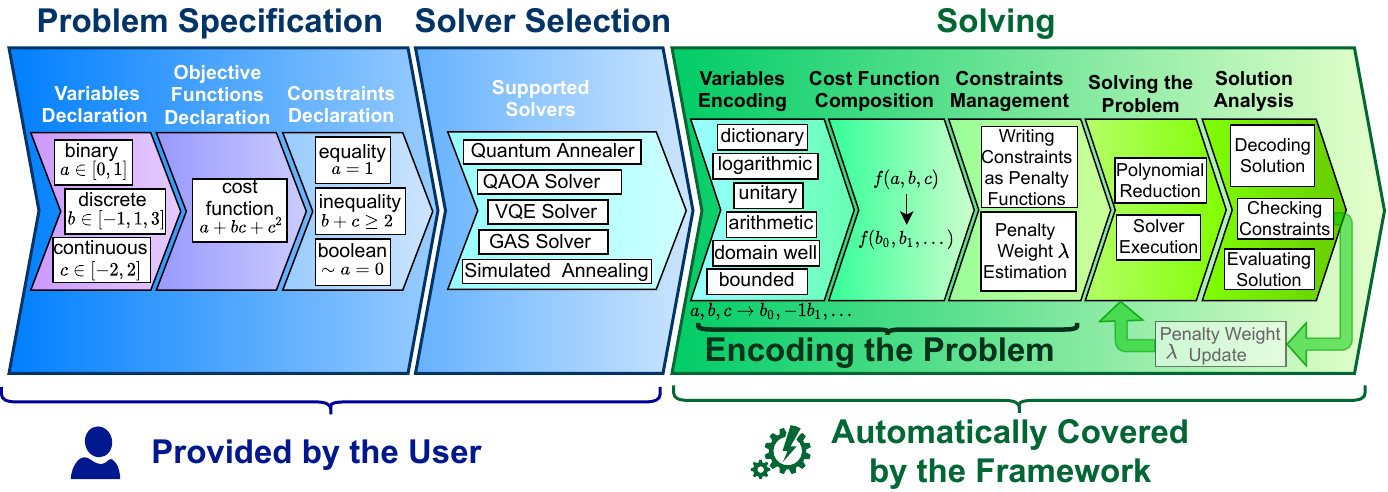}
\caption{Quantum Optimization Flow. \vspace{-15pt}}
\label{fig:ASS}
\end{figure*}
\vspace{-3pt}
\section{Towards an Automatic Framework for Solving Optimization Problems}\label{sec:Towards}
\vspace{-3pt}
This section presents the unmet needs that the proposed framework seeks to address,  discusses the idea behind it, and offers a \mbox{high-level} description of its structure. Additionally, it compares the framework with existing \mbox{state-of-the-art libraries and tools}. 
\vspace{-5pt}
\subsection{Motivations}
\vspace{-3pt}
All the steps reviewed above demand significant expertise in QUBO formulation and quantum solvers, which is uncommon among conventional optimization users. This expertise gap limits accessibility to the potential quantum computational advantage. Moreover, some tasks, such as variable encoding, solution decoding, as well as function rewriting can be cumbersome, and \textit{manual execution} may introduce \textit{errors}. Managing penalty functions is also critical, especially for translating constraints, which is not always effortless, especially when \textit{auxiliary variables} are needed, e.g., for inequality constraints. Furthermore, determining an \textit{appropriate penalty weight} is crucial for obtaining good results. A too-low value may not adequately penalize invalid solutions, while a too-high value may flatten the optimization function, complicating effective solution space exploration. Lastly, the configuration and the execution of solvers might not be intuitive for non-experts in quantum computing. For all these reasons, these steps can greatly benefit from \textit{automation} for both \textit{preventing errors} arising from the manual translation of the optimization problem into a solver-compatible format and \textit{improving accessibility} to quantum solvers for a wider range of users. 

\subsection{General Idea}
This work aims to offer a versatile and \textit{\mbox{user-friendly}} framework for assisting non-experts in quantum solvers exploitation for \mbox{real-world problems}, keeping the user in the conventional optimization domain by requiring as input the problem description in its simplest and most intuitive formulation. In particular, the user interface and the required inputs are analogous to one of \textit{Gurobi}~\cite{gurobipy-documentation}, 
 a well-consolidated optimization package for conventional solvers.

The proposed framework aims to assist non-experts in exploring the potential of the quantum solver in the same fashion as classical alternatives. It enables the automated translation of a problem from a classical description to a format compatible with the chosen quantum solver. Moreover, it allows and aids in decoding and evaluating the obtained solution. 

Figure \ref{fig:ASS} summarizes the corresponding steps to be done by the user and those that are automatically covered by the proposed framework. More precisely:
\begin{enumerate}
\item The user must provide the \textit{problem specification}, consisting of \textit{variables declaration}, which can be binary (unipolar or bipolar), discrete (specifying the valid values), and continuous (defining the operative range and the wanted precision), \textit{objective functions assertion}, specifying the importance weight, if a multi-objective optimization occurs, as well as
\textit{constraints definition}, belonging to equality, inequality, and boolean categories.
\item Then, the user has to \textit{select a solver} among the supported ones, i.e., QA, QAOA, VQE, GAS, or SA; 
\item Based on that, the framework \textit{automatically} translates the description into a solver-compatible format and analyzes the obtained solution by executing the following steps.  
\begin{enumerate}
		\item \textit{Variables encoding}, which consists of describing the multilevel variables as a set of binary ones with one of the techniques described in Section \ref{sec:variablesEncoding}.
		\item \textit{Composing the cost function} involving only binary variables, as Section \ref{sec:CostFunctionComposition} describes.
		\item \textit{Writing the constraints as penalty functions}, and \textit{estimating a proper penalty weight} $\lambda$, as described in Sections \ref{sec:ProblemConstraint} and \ref{sec:lambdadef}, respectively.
               \item \textit{Solving the problem} with the chosen optimizer, as described in Section \ref{sec:Solvers}.
		\item \textit{Solution analysis}, checking the constraint satisfaction, and eventually \textit{updating the penalty weight} $\lambda$ for obtaining valid outcomes, as described in Sections \ref{sec:SolutionAnalysis} and \ref{sec:lambdaupdate}.
	\end{enumerate}
\end{enumerate}


{\renewcommand{\arraystretch}{1.5}
\setlength{\tabcolsep}{2pt}
\begin{table*}
\caption{Comparing the support provided by of proposed framework and existing libraries and framework in each step of quantum optimization.\\ \textcolor{ForestGreen}{\cmark} indicates that the corresponding action is performed automatically.\\ \textcolor{YellowOrange}{\cmark} signifies that a proper function is available for implementing the step. \\  \textcolor{red}{\xmark} indicates that the method  is not fully supported. \\ $^+$ denotes that logarithmic encoding is also compatible with bases different from two.\\ $^*$ signifies that the encoding techniques can be exploited only for constraints translation.\\
$^{\dagger}$ indicates that the polynomial reduction is implemented by exploiting the corresponding qubovert function.
 }
\begin{center}
\resizebox{0.982415\textwidth}{!}{
\begin{tabular}{?c|c?c|c|c|c|c|c?c|c?c?}
\noalign{\hrule height 1.5pt}
\multicolumn{2}{?c?}{\multirow{2}{*}{\textbf{Supports for each step}}} & \multicolumn{6}{c?}{\textbf{ Existing Libraries}}& \multicolumn{2}{c?}{\textbf{Existing Frameworks}} &\textbf{Proposed}\\
\cline{3-10}
\multicolumn{2}{?c?}{}& \textbf{pyqubo\cite{pyqubo-docs}} & \textbf{qubovert\cite{qubovert-docs}} & \textbf{dimod\cite{dimod-docs}} & \textbf{Qiskit\cite{qiskit-optimization-docs}} & \textbf{fixstars 
\cite{fixstarts-docs}} &\textbf{openQAOA\cite{openQAOA-entropica-labs-docs}} & \textbf{AutoQUBO \cite{autoqubo-repository}} & \textbf{QUBO.jl \cite{QUBO.jl-repository}} & \textbf{Framework}\\
\noalign{\hrule height 1.5pt}
\multicolumn{2}{?c?}{\textbf{Floating Encoding}} & \textcolor{red}{\xmark} &  \textcolor{red}{\xmark} &  \textcolor{red}{\xmark} &  \textcolor{red}{\xmark} &  \textcolor{red}{\xmark} & \textcolor{red}{\xmark} &  \textcolor{red}{\xmark} &   \textcolor{ForestGreen}{\cmark} & \textcolor{ForestGreen}{\cmark}\\
 \hline 
 \multirow{5}{*}{\textbf{Integer Encoding}} & \textbf{Logarithmic \cite{tamura2021performance}} & \textcolor{YellowOrange}{\cmark} & \textcolor{YellowOrange}{\cmark} & \textcolor{YellowOrange}{\cmark} & \textcolor{YellowOrange}{\cmark} & \textcolor{YellowOrange}{\cmark}$^*$ & 
 \textcolor{red}{\xmark} & \textcolor{red}{\xmark} & \textcolor{ForestGreen}{\cmark} & \textcolor{ForestGreen}{\cmark}$^+$\\
 \cline{2-11}
 & \textbf{Unitary\cite{tamura2021performance}} & \textcolor{YellowOrange}{\cmark}  & \textcolor{YellowOrange}{\cmark}  &  \textcolor{red}{\xmark} &  \textcolor{red}{\xmark} &  \textcolor{YellowOrange}{\cmark}$^*$  & \textcolor{red}{\xmark}&  \textcolor{red}{\xmark} & \textcolor{ForestGreen}{\cmark} &\textcolor{ForestGreen}{\cmark}\\
  \cline{2-11}
 & \textbf{Dictionary \cite{tamura2021performance}} & \textcolor{YellowOrange}{\cmark} &  \textcolor{red}{\xmark} &  \textcolor{red}{\xmark} & \textcolor{red}{\xmark} &  \textcolor{red}{\xmark} & \textcolor{red}{\xmark} &  \textcolor{red}{\xmark} & \textcolor{ForestGreen}{\cmark} & \textcolor{ForestGreen}{\cmark}\\
   \cline{2-11}
 & \textbf{Domain-Wall \cite{chancellor2019domain}} & \textcolor{YellowOrange}{\cmark} & \textcolor{red}{\xmark} & \textcolor{red}{\xmark} & \textcolor{red}{\xmark} & \textcolor{red}{\xmark} &\textcolor{red}{\xmark} & \textcolor{red}{\xmark} & \textcolor{ForestGreen}{\cmark} & \textcolor{ForestGreen}{\cmark}\\
    \cline{2-11}
 & \textbf{Bounded-Coeff \cite{karimi2019practical}} & \textcolor{red}{\xmark} & \textcolor{red}{\xmark} & \textcolor{red}{\xmark} & \textcolor{red}{\xmark} & \textcolor{red}{\xmark} &\textcolor{red}{\xmark} & \textcolor{red}{\xmark} & \textcolor{ForestGreen}{\cmark}& \textcolor{ForestGreen}{\cmark}\\
     \cline{2-11}
  & \textbf{Arithmetic \cite{xavier2023qubo}} & \textcolor{red}{\xmark}  & \textcolor{red}{\xmark} & \textcolor{red}{\xmark}  & \textcolor{red}{\xmark}  & \textcolor{YellowOrange}{\cmark}$^*$ & \textcolor{red}{\xmark}  & \textcolor{red}{\xmark}  & \textcolor{ForestGreen}{\cmark} & \textcolor{ForestGreen}{\cmark}\\
  \hline
   \multirow{3}{*}{\textbf{Penalty Functions}} & \textbf{Equality \cite{combarro2023practical}} \cite{glover2018tutorial} &  \textcolor{red}{\xmark} & \textcolor{YellowOrange}{\cmark} & \textcolor{YellowOrange}{\cmark} & \textcolor{YellowOrange}{\cmark} & \textcolor{YellowOrange}{\cmark} &\textcolor{YellowOrange}{\cmark} &  \textcolor{red}{\xmark} & \textcolor{ForestGreen}{\cmark} &\textcolor{ForestGreen}{\cmark}\\
   \cline{2-11}
   & \textbf{Inequality \cite{combarro2023practical}} &  \textcolor{red}{\xmark} & \textcolor{YellowOrange}{\cmark} & \textcolor{YellowOrange}{\cmark} & \textcolor{YellowOrange}{\cmark} & \textcolor{YellowOrange}{\cmark} &\textcolor{YellowOrange}{\cmark} &  \textcolor{red}{\xmark} &\textcolor{ForestGreen}{\cmark} & \textcolor{ForestGreen}{\cmark}\\
      \cline{2-11}
   & \textbf{Boolean \cite{combarro2023practical}} & \textcolor{YellowOrange}{\cmark} & \textcolor{YellowOrange}{\cmark} &\textcolor{YellowOrange}{\cmark} &  \textcolor{red}{\xmark} & \textcolor{YellowOrange}{\cmark} & \textcolor{red}{\xmark}& \textcolor{red}{\xmark} &  \textcolor{red}{\xmark} &\textcolor{ForestGreen}{\cmark}\\
   \hline
   \multirow{7}{*}{\textbf{Penalty Weight}} & \textbf{UB positive \cite{ayodele2022penalty}}&  \textcolor{red}{\xmark} &  \textcolor{red}{\xmark} &  \textcolor{red}{\xmark} &  \textcolor{red}{\xmark} &  \textcolor{red}{\xmark} & \textcolor{red}{\xmark} &  \textcolor{red}{\xmark} &  \textcolor{red}{\xmark}& \textcolor{ForestGreen}{\cmark}\\
   \cline{2-11}
   & \textbf{MQC \cite{ayodele2022penalty}}&  \textcolor{red}{\xmark} &  \textcolor{red}{\xmark} & \textcolor{YellowOrange}{\cmark} &  \textcolor{red}{\xmark} &  \textcolor{red}{\xmark} & \textcolor{red}{\xmark} &  \textcolor{red}{\xmark} &  \textcolor{red}{\xmark} & \textcolor{ForestGreen}{\cmark}\\
   \cline{2-11}
      & \textbf{VLM \cite{verma2022penalty}}&  \textcolor{red}{\xmark} &  \textcolor{red}{\xmark} &  \textcolor{red}{\xmark} &  \textcolor{red}{\xmark} &  \textcolor{red}{\xmark} & \textcolor{red}{\xmark} & \textcolor{ForestGreen}{\cmark} &  \textcolor{red}{\xmark}& \textcolor{ForestGreen}{\cmark}\\
   \cline{2-11}
         & \textbf{MOMC \cite{ayodele2022penalty}}&  \textcolor{red}{\xmark} &  \textcolor{red}{\xmark} & \textcolor{red}{\xmark} & \textcolor{red}{\xmark} &  \textcolor{red}{\xmark} & \textcolor{red}{\xmark} &  \textcolor{red}{\xmark} &  \textcolor{red}{\xmark} & \textcolor{ForestGreen}{\cmark}\\
   \cline{2-11}
            & \textbf{MOC \cite{ayodele2022penalty}}&  \textcolor{red}{\xmark} &  \textcolor{red}{\xmark} &  \textcolor{red}{\xmark} &  \textcolor{red}{\xmark} &  \textcolor{red}{\xmark} & \textcolor{red}{\xmark} &  \textcolor{red}{\xmark} &  \textcolor{red}{\xmark} & \textcolor{ForestGreen}{\cmark}\\
   \cline{2-11}
            & \textbf{UB Naive \cite{boros2002pseudo, boros2006preprocessing}}&  \textcolor{red}{\xmark} & \textcolor{YellowOrange}{\cmark} &  \textcolor{red}{\xmark} & \textcolor{ForestGreen}{\cmark} &  \textcolor{red}{\xmark} & \textcolor{red}{\xmark} & \textcolor{ForestGreen}{\cmark} &  \textcolor{red}{\xmark} & \textcolor{ForestGreen}{\cmark}\\
   \cline{2-11}
               & \textbf{UB posiform \cite{boros2002pseudo, boros2006preprocessing}}&  \textcolor{red}{\xmark} &  \textcolor{red}{\xmark} &  \textcolor{red}{\xmark} &  \textcolor{red}{\xmark} &  \textcolor{red}{\xmark} & \textcolor{red}{\xmark} & \textcolor{ForestGreen}{\cmark} &  \textcolor{red}{\xmark} & \textcolor{ForestGreen}{\cmark}\\

          \hline
   \multicolumn{2}{?c?}{\textbf{Polynomial Reduction}} & \textcolor{YellowOrange}{\cmark} & \textcolor{YellowOrange}{\cmark} & \textcolor{YellowOrange}{\cmark} & \textcolor{red}{\xmark} & \textcolor{YellowOrange}{\cmark} & \textcolor{red}{\xmark} & \textcolor{ForestGreen}{\cmark} & \textcolor{ForestGreen}{\cmark} &\textcolor{ForestGreen}{\cmark}$^{\dagger}$\\
     \hline
    \multirow{5}{*}{\textbf{Solvers}} &  \textbf{Dwave QA} & \textcolor{ForestGreen}{\cmark} & \textcolor{red}{\xmark} & \textcolor{ForestGreen}{\cmark} & \textcolor{red}{\xmark} & \textcolor{ForestGreen}{\cmark} &\textcolor{red}{\xmark} & \textcolor{ForestGreen}{\cmark} & \textcolor{ForestGreen}{\cmark} & \textcolor{ForestGreen}{\cmark}\\
    \cline{2-11}
     & \textbf{QAOA} &\textcolor{red}{\xmark} & \textcolor{red}{\xmark} & \textcolor{red}{\xmark} & \textcolor{ForestGreen}{\cmark} & \textcolor{red}{\xmark} &\textcolor{ForestGreen}{\cmark} & \textcolor{red}{\xmark} & \textcolor{ForestGreen}{\cmark} & \textcolor{ForestGreen}{\cmark}\\
    \cline{2-11}
 & \textbf{VQE} & \textcolor{red}{\xmark} & \textcolor{red}{\xmark} & \textcolor{red}{\xmark} & \textcolor{ForestGreen}{\cmark} & \textcolor{red}{\xmark} &\textcolor{ForestGreen}{\cmark} & \textcolor{red}{\xmark} & \textcolor{ForestGreen}{\cmark} & \textcolor{ForestGreen}{\cmark}\\
     \cline{2-11}
 & \textbf{GAS} & \textcolor{red}{\xmark} & \textcolor{red}{\xmark} & \textcolor{red}{\xmark}& \textcolor{ForestGreen}{\cmark} & \textcolor{red}{\xmark} &\textcolor{red}{\xmark} & \textcolor{red}{\xmark} & \textcolor{red}{\xmark} & \textcolor{ForestGreen}{\cmark}\\
\cline{2-11}
 & \textbf{SA} &\textcolor{ForestGreen}{\cmark}& \textcolor{ForestGreen}{\cmark} & \textcolor{ForestGreen}{\cmark} & \textcolor{ForestGreen}{\cmark} & \textcolor{ForestGreen}{\cmark} &\textcolor{red}{\xmark} & \textcolor{ForestGreen}{\cmark} & \textcolor{ForestGreen}{\cmark} & \textcolor{ForestGreen}{\cmark}\\

       \hline
      \multicolumn{2}{?c?}{\textbf{Solution Decoding}} & \textcolor{YellowOrange}{\cmark} & \textcolor{YellowOrange}{\cmark} & \textcolor{YellowOrange}{\cmark}& \textcolor{YellowOrange}{\cmark} & \textcolor{YellowOrange}{\cmark} & \textcolor{YellowOrange}{\cmark} & \textcolor{ForestGreen}{\cmark}& \textcolor{ForestGreen}{\cmark} & \textcolor{ForestGreen}{\cmark}\\
   \hline
      \multicolumn{2}{?c?}{\textbf{Check Constraints}} & \textcolor{YellowOrange}{\cmark} & \textcolor{YellowOrange}{\cmark} & \textcolor{YellowOrange}{\cmark} & \textcolor{red}{\xmark} & \textcolor{YellowOrange}{\cmark} & \textcolor{red}{\xmark} & \textcolor{ForestGreen}{\cmark} & \textcolor{ForestGreen}{\cmark} & \textcolor{ForestGreen}{\cmark}\\

       \hline

   \multirow{3}{*}{\textbf{Penalty Update}} & \textbf{Sequential \cite{garcia2022exact}}& \textcolor{red}{\xmark} & \textcolor{red}{\xmark} & \textcolor{red}{\xmark} & \textcolor{red}{\xmark} & \textcolor{red}{\xmark} & \textcolor{red}{\xmark} & \textcolor{red}{\xmark} & \textcolor{red}{\xmark} & \textcolor{ForestGreen}{\cmark}\\
   \cline{2-11}
     & \textbf{Scaled \cite{garcia2022exact}}& \textcolor{red}{\xmark} & \textcolor{red}{\xmark} & \textcolor{red}{\xmark} & \textcolor{red}{\xmark} & \textcolor{red}{\xmark} &\textcolor{red}{\xmark} & \textcolor{red}{\xmark} & \textcolor{red}{\xmark} & \textcolor{ForestGreen}{\cmark}\\
   \cline{2-11}
        & \textbf{Binary search \cite{garcia2022exact}}& \textcolor{red}{\xmark} & \textcolor{red}{\xmark} & \textcolor{red}{\xmark} & \textcolor{red}{\xmark} & \textcolor{red}{\xmark} &\textcolor{red}{\xmark} & \textcolor{red}{\xmark} & \textcolor{red}{\xmark} & \textcolor{ForestGreen}{\cmark}\\

\noalign{\hrule height 1.5pt}
\hline
\end{tabular} }\vspace{-20pt}
\end{center}
\label{tab:TableComparisonsWithLibrary}
\end{table*}

\subsection{Related Works}

In recent years, supporting and automating the procedure for describing optimization problems in a \mbox{quantum-compliant} format for quantum solver exploitation has become a focus. Several libraries and some tools have emerged to aid the QUBO formulation process. The main libraries include \textit{pyqubo}~\cite{pyqubo-docs, zaman2021pyqubo}, \textit{qubovert}~\cite{qubovert-docs}, \textit{dimod}~\cite{dimod-docs}, \textit{Qiskit-optimization}~\cite{qiskit-optimization-docs}, \textit{Fixstarts Amplify}~\cite{fixstarts-docs} and \textit{openQAOA Entropica}~\cite{openQAOA-entropica-labs-docs}, while two frameworks, \textit{AutoQUBO}~\cite{autoqubo-repository, moraglio2022autoqubo,pauckert2023autoqubo} and QUBO.jl~\cite{xavier2023qubo}, have been proposed in the last two years to meet users' requests for a tool that automates the \textit{entire} procedure. Table~\ref{tab:TableComparisonsWithLibrary}  provides a comprehensive overview of all these tools together with a comparison of the features of the framework proposed in this work aims to provide.

More precisely, the table shows that while libraries significantly simplify complex steps in their respective procedures, their main limitation lies in the lack of support for the automatic execution of these steps, restricting their usability to users with at least a minimum level of expertise in the field. Moreover, each library alone does not cover all essential steps completely. On the other hand,  the main limitation of AutoQUBO is its lack of support for managing non-binary variables and writing constraints as penalty functions. Furthermore, it is principally designed for \textit{Digital Annealer} (DA)~\cite{aramon2019physics}---the quantum-inspired Fujitsu solver---, the SA and the QA, with no support for execution on quantum-circuit-model-based solvers. Finally,  QUBO.jl~\cite{xavier2023qubo}  requires as input a JuMP problem (expressed in the Julia language instead of Python, which is the main language for quantum frameworks) and stands out as the first tool supporting the encoding of floating variables. However, it is incomplete concerning the penalties weight assignment. 

As can be seen, all existing libraries and frameworks still heavily rely on the end user and do not fully harness the potential for automation. In contrast, to the best of our knowledge, the framework proposed in this work stands out as \textit{the first to guarantee complete coverage of the most crucial steps} of quantum optimization. Moreover, it supports all methods available in the literature for each step, allowing the selection of the most suitable approach for a specific application. 


\begin{figure*}[t]
	\centering
	\includegraphics[width=1\textwidth]{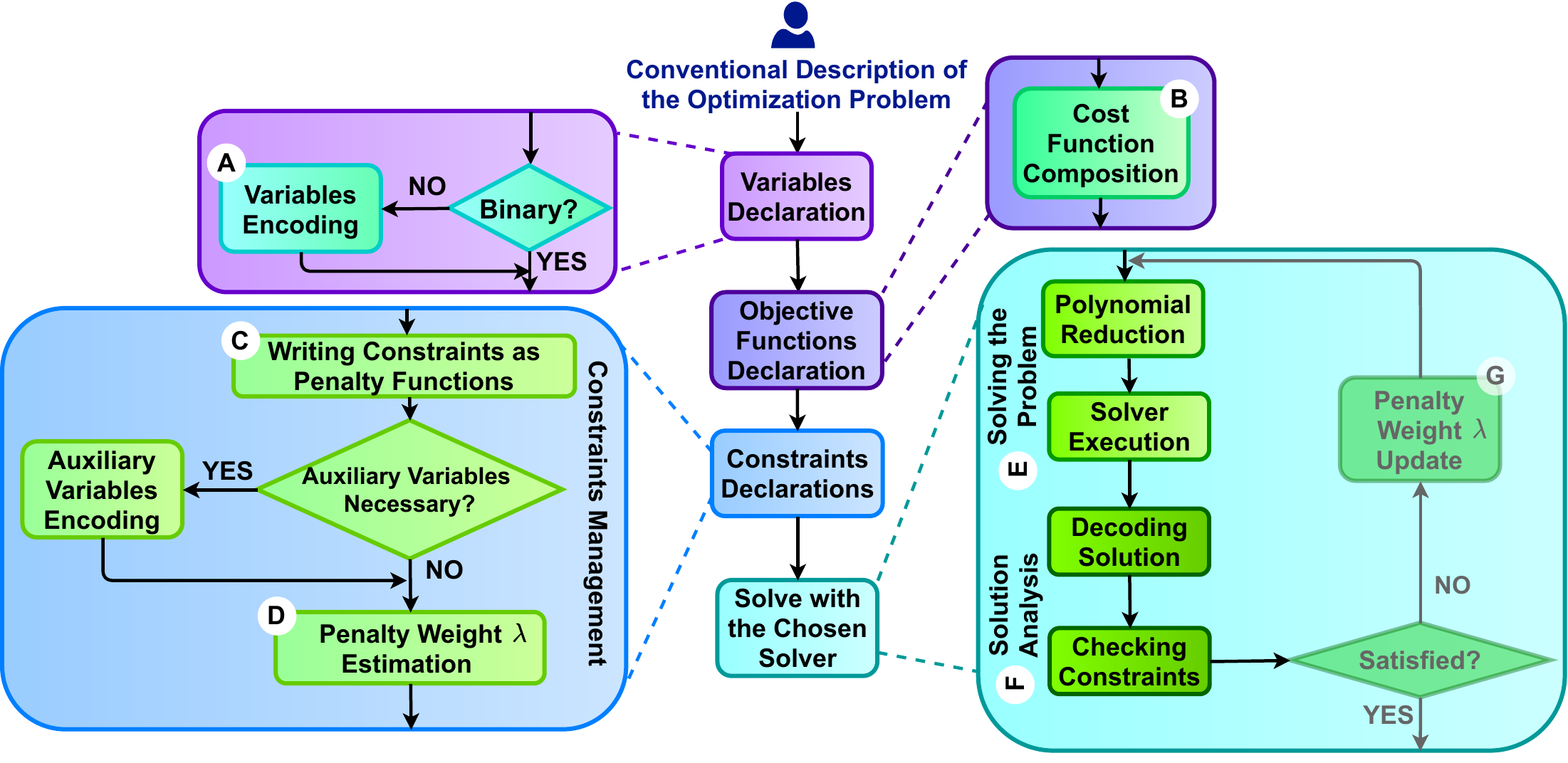}
\caption{Workflow of the proposed framework. The letters correspond to the subsections where each step is discussed. \vspace{-10pt}}
\label{fig:QOT}
\end{figure*}
\vspace{-3pt}
\section{Implementation of the Proposed Framework} \label{sec:Implementation}
\vspace{-3pt}
This section details the implementation of the proposed framework, designed to automate the optimization of any polynomial cost functions exploiting quantum solvers. Figure~\ref{fig:QOT} shows the workflow considered.  The example introduced in the following is examined to adequately describe the steps required for transitioning a generic cost function into a quantum-solver-compliant format. 
\vspace{2pt}
\examplethree{Let us consider the cost function $a+ b c + c^2$, where $a$ is a unipolar binary variable, $b$ is a discrete variable taking values in the set $[-1, 1, 3]$, and $c$ is a continuous variable in the range $[-2,2]$. The target is to determine the configuration of variables that minimizes the cost function while satisfying the constraint 
 $b+c \geq 2$. }
\vspace{3pt}
This abstract example has been chosen because its simplicity enables us to effectively showcase various aspects of the framework.
Using this example, the following describes how the steps illustrated in Figure \ref{fig:QOT} are automatically \mbox{conducted}.  

 \vspace{-3pt}
\subsection{Variable Encoding} \label{sec:variablesEncoding}
\vspace{-3pt}
First, the variables must be converted into a binary format through proper encoding techniques, whose target is to associate with each multilevel variable a set of \textit{binary} ones with \textit{appropriate weights} and, eventually, \textit{offset}. The employed encoding technique depends on the type of variable declared by the user. In particular, for \textit{unipolar binary} variables, a \mbox{one-to-one} association is established without the need for weights or offsets; for \textit{bipolar binary} variables, a \mbox{one-to-one} association is considered with a weight of two and an offset of minus one; for \textit{discrete} variables, a \textit{dictionary} encoding~\cite{tamura2021performance} mechanism can be implemented by associating a binary variable with each possible value, ensuring that only one of them can be active in a valid solution; for \textit{continuous} variables, various encoding techniques, for a wanted precision, are supported such as \textit{dictionary} encoding~\cite{tamura2021performance} (as in the discrete case exploiting operative range discretization), \textit{logarithmic} encoding~\cite{tamura2021performance}, \textit{unitary} encoding~\cite{tamura2021performance}, \textit{arithmetic progression} encoding~\cite{xavier2023qubo}, \mbox{\textit{domain-well}} encoding~\cite{chancellor2019domain}, and \textit{bounded coefficient} encoding~\cite{karimi2019practical}.

In the proposed framework, an effort was made to extend the mentioned encoding techniques,  initially designed for \textit{integers}, to \textit{real} numbers, even managing an asymmetric operational range. Moreover, the framework allows the declaration of multidimensional arrays of variables. This feature enables a more concise definition of objective functions, allowing the direct expression of matrix-based formulations.
\examplethree{The presented variables can be encoded in the following way:}
\begin{itemize}
    \item $a \rightarrow b_0$;
    \item $b \rightarrow -1b_1+1b_2+3b_3$, \textit{imposing} $b_1+b_2+b_3 = 1$;
    \item $c \rightarrow$:
    \begin{itemize}
        \item \textit{dictionary considering a precision of 0.5:} $-2b_4-1.5b_5-1b_6-0.5b_7+0b_8+0.5b_9+1b_{10}+1.5b_{11}+ 2b_{12}$, \textit{imposing} $b_4+b_5+b_6+b_7+b_8+b_9+b_{10}+b_{11}+ b_{12} = 1$;
        \item \textit{logarithmic considering a precision of 0.25 and the base 2:} $0.25b_4 + 0.5b_5+1b_6+2b_7+0.25b_8 -2$;
        \item \textit{unitary considering a precision of 0.5:} $0.5b_4+0.5b_5+0.5b_6+0.5b_7+0.5b_8+0.5b_9+0.5b_10+0.5b_{11}-2$;
        \item \textit{arithmetic progression considering a precision of 0.2:} $0.2b_4+0.4b_5+0.6b_6+0.8b_7+1b_8+1b_9-2$;
        \item \textit{domain-well considering a precision of 0.5:} $0.5b_4+1b_5+1.5b_6+2b_7+2.5b_8+2b_9-2$ \textit{imposing} $b_i \geq b_{i-1}$;
        \item \textit{bounded coefficient considering a precision of 0.5, with bound 1:} $0.5b_4+1b_5+1b_6+1b_7-2$.
    \end{itemize}
\end{itemize}

\vspace{-3pt}
\subsection{Cost Function Composition} \label{sec:CostFunctionComposition}
\vspace{-3pt}
Afterwards, the declared objective functions have to be combined, considering the \textit{aggregation weight}, the optimization sense (adjusting the sign for maximization contributions), and replacing the original variables with the binary ones, as in the following:
\begin{equation}
\vspace{-4pt}
    (\min, f(x), \delta), (\max, g(y), \eta) \rightarrow q(\textbf{b}) = \delta f(\textbf{b}) - \eta g(\textbf{b}) \, , 
\end{equation}
where $x$ and $y$ are the variables declared by the user, $\textbf{b}$ the array of binary variables for the encoding, $\min$ and $\max$ indicate minimization and maximization, respectively, $\delta$ and $\eta$ are the user preference weights. 
The result is a cost function considering all the optimization figures of merit involving exclusively binary variables. 

\examplethree{Considering the logarithmic encoding mechanism with a precision of 0.25 and the base 2 for variable $c$, the declared objective function $a + bc + c^2$ becomes:
\begin{align*}
       &a + bc + c^2 \rightarrow (b_0) + (-1b_1 +1 b_2 + 3b_3)(0.25b_4 +\\&+ 0.5b_5+1b_6+2b_7+ 0.25b_8 -2)
       + (0.25b_4 + 0.5b_5+\\ &+1b_6+2b_7+0.25b_8 -2)^2 
\end{align*} 
\vspace{-2pt}
Finally, the expression is expanded.}
\subsection{Writing Constraints as Penalty Functions}\label{sec:ProblemConstraint}
\vspace{-3pt}
Subsequently, the problem constraints are transformed into appropriate penalty functions within the framework. This step consists of identifying the type of constraint, replacing the original variable with binary counterparts in the constraint expression, and composing the proper penalty function associated with the constraint, eventually inserting auxiliary variables. 

The framework can handle \textit{equality constraints}~\cite{combarro2023practical},  \textit{inequality constraints} (greater than, greater equal, less than, and less equal)---expressed as equality constraints with the introduction of auxiliary variables ~\cite{combarro2023practical}, whose necessary precision can be defined by the user or inferred by the expression---, and \textit{boolean constraints} (not, and, or, xor).  These constraint categories aim to ensure comprehensive coverage of prevalent situations encountered in real-world scenarios.
\vspace{3pt}
\examplethree{The first constraint to evaluate in the proposed example is one deriving from the encoding of variable $b$, i.e., $b_1+b_2+b_3 = 1$. Therefore, the proper penalty function expressing this constraint is $g(\textbf{b}) = (b_1+b_2+b_3 - 1)^2$. In this case, the constraint is already written with the final problem binary variables, not requiring the substitution step.}
\noindent\textit{The expression can be expanded as:
\begin{equation*}
    g(\textbf{b}) = - b_1 - b_2 - b_3 + 2b_1b_2 + 2b_1b_3 + 2b_2b_3 + 1 \, ,
\end{equation*}
remembering that $b_i = b_i^2$. It is possible to notice, that, as desired, $g(x)$ is null only when exactly one among the variables is set to one and assumes a positive value otherwise.   \\
The second constraint is imposed by the user and requires that $ b + c  \geq 2$. First of all, this has to be written as an equality constraint through the introduction of a real auxiliary variable $\_\_aux$, which should have the operative range $[-(\max(b+c)-2),0]$, i.e.,  $[-3,0]$. In this way, the constraint becomes $ b + c + \_\_aux = 2$, which can be written as $k(b,c) = (b + c + \_\_aux - 2)^2$. It is possible to notice that, choosing proper values of the auxiliary variable, all the $b,c$ configurations satisfying the constraint are not penalized (e.g., for $b=3$, $c=1$, the penalty function is null for $\_\_aux =-2$). Then, the variables are replaced in the function with their binary encoding (logarithm is considered for the auxiliary variable) as follows:
\begin{align*}
    &k(\textbf{b}) = ( 0.25\_a_0 + 0.5\_a_1 + 1\_a_2 + 1.25\_a_3 -1b_1 +1b_2  +\\&3b_3 +0.25b_4 + 0.5b_5 + 1b_6 + 2b_7 +0.25b_8 -2  -3 )^2\, .
\end{align*}
Finally, the expression is expanded. }

\subsection{Penalty Weight Estimation}\label{sec:lambdadef}
Determining the \textit{penalty weight}, denoted as $\lambda$,  is critical for \textit{effectively penalizing non-valid solutions} while integrating penalty functions with the objective function. In particular, it is necessary to select a $\lambda$ satisfying the following relation: 
\begin{equation*}
    f(x*) < f(x) + \lambda g(x) \quad \forall x \in S \rightarrow \lambda > \max_{x \in S}\biggl( \frac{f(x*) - f(x)}{g(x)}\biggr) \, 
\end{equation*}
where $f(x)$ is the cost function to optimize, $f(x*)$ is the cost function value for the  optimal solution $x*$, $g(x)$ is the penalty function and $S$ is the space of infeasible solutions. Accordingly, its evaluation would require knowing the range of values that cost and penalty functions assume  (functions bounds).  Therefore, \textit{limited complexity methods} for estimating a $\lambda$ satisfying the relation, i.e.,  ensuring the correct penalization of unfeasible solutions without compromising the quality of the solution space exploration. The state-of-the-art suggests taking $\lambda$ equal to a certain percentage of the objective function range, i.e., about \textit{75\%-150\%}~\cite{glover2018tutorial}. 

The framework supports the following methods of literature: 
\begin{itemize}
    \item \textit{Upper bound of objective functions involving only positive coefficients} (UB positive, ~\cite{ayodele2022penalty}), which is valid for QUBO problems with all positive coefficients and estimates the objective function upper bound by summing all the QUBO coefficients.
    \item \textit{Maximum QUBO coefficient} (MQC), which is originally thought for travelling salesman problem and chooses as $\lambda$ the QUBO coefficient presenting the highest value.
    \item \textit{Verma and Lewis} (VLM, ~\cite{verma2022penalty}), which estimates the amount of gain/loss that an objective function can achieve by flipping a single variable. $\lambda$ is its maximum value.
    \item \textit{Maximum change in Objective function divided by Minimum Constraint function of infeasible solutions} (MOMC, ~\cite{ayodele2022penalty}), which is a modified version of VLM, dividing its $\lambda$, by the minimum variation in the constraint function.
    \item \textit{Maximum value derived from dividing each change in Objective function with the corresponding change in Constraint function} (MOC, ~\cite{ayodele2022penalty}), which is derived from VLM, dividing each change in the objective function by the corresponding change in the constraint function and choosing the maximum ratio.
    \item \textit{Upper bound Naive} (UB Naive, ~\cite{boros2002pseudo, boros2006preprocessing}), which estimates the cost function upper bound as the sum of all positive QUBO coefficients and the lower bound as the sum of negatives. The $\lambda$ is the difference between the upper and lower bound. 
    \item \textit{Upper Bound posiform} (UB posiform, ~\cite{boros2002pseudo, boros2006preprocessing}), which computes upper and lower bounds by exploiting the concept of posiform and negaform.
\end{itemize}
It is possible to notice that MOMC and MOC are unique policies considering not only the objective functions but also the penalties to apply.

In the proposed framework, the values found with the presented method are multiplied by an amount, close to one, which can be chosen by the user, which is different for hard and weak constraints.  This provides an additional degree of freedom for the user to adapt the approaches to their needs.  Alternatively, the $\lambda$'s values can be set manually, although is not recommended.  \\
\examplethree{For the considered example, the values found for $\lambda$ are as follows: 
\begin{itemize}
    \item UB positive: Not applicable;
    \item MQC: $\lambda = 6 + 4 = 10$, where $6$ is the maximum coefficient and $4$ the function offset, for both constraints;
    \item VLM: $\lambda=12$, for both constraints;
    \item MOMC:  $\lambda_0=6.19$ and  $\lambda_1=12$, for the inequality and equality constraints, respectively;
    \item MOC: $\lambda_0=1$ and $\lambda_1=6$, for the inequality and equality constraints, respectively;
    \item UB Naive: $\lambda=52.25$, for both constraints 
    \item UB posiform: $\lambda=31.625$, for both constraints.
\end{itemize}}

\subsection{Solving the Problem}\label{sec:Solvers}
Next, the problem is optimized by exploiting the selected solver. However, a preliminary step involves \textit{reducing polynomials to second order} before proceeding. In our framework, this task is currently performed by functions within the qubovert library. This choice was made for simplicity, as the method for this reduction is essentially standard, with few degrees of freedom and possible optimizations with respect to the other steps of the tool, and was already well-implemented in the library. Unfortunately, polynomial reduction is expensive from the problem complexity point of view, as it involves adding \textit{auxiliary binary variables} to represent polynomials with degrees higher than two.

The framework offers a \textit{common interface} for all supported solvers, with solutions provided in the same format, as discussed in the next section. However, setting optimizer parameters differs between solvers, as they are specific to each solver's requirements.\\

\vspace{-10pt}
\subsection{Solution Analysis}\label{sec:SolutionAnalysis}
\vspace{-3pt}
The solvers provide in output a solution object comprising: 
\begin{itemize}
\item the list of obtained \textit{solutions} in binary variables;
\item the list of obtained \textit{solutions} in the originally  declared variables through a proper \textit{decoding} mechanism;
\item list of \textit{energies} found;
\item the \textit{best solution} in binary and originally declared variables, i.e., ones corresponding to the  lowest energy; 
\item the \textit{best energy} value.
\end{itemize}
Moreover, the framework provides a set of methods for analyzing the obtained results:
\begin{itemize}
\item Methods for \textit{checking constraint satisfaction} of the best solution or all the obtained solutions, eventually considering weak constraints if a flag is set. 
\item Method for evaluating the \textit{value} obtained with the best solution \textit{for each objective function} declared (particularly useful in case of multi-objective optimization).
\item Method for showing the \textit{cumulative distribution} of the obtained results. To understand the meaning of this, one rule has to be considered: the probability of obtaining the optimal value (or a value close to it) with a solver is higher when its corresponding cumulative distribution is more concentrated on the left of the plot, where the lowest values are located. 
\item Method for evaluating the \textit{rate of valid solutions} found. 
\item Method for evaluating the $p_\textrm{range}$, which is the probability of obtaining a final energy lower than a certain value (val\_ref) and can be computed as follows:
\begin{equation}
\vspace{-2pt}
    p_{\textrm{range}} \triangleq \frac{n_{\textrm{in\_range}}}{n_{\textrm{tot}}} 100 \, ,
    \label{eq:p_range}
\vspace{-1pt}
\end{equation}
where $n_{\textrm{in\_range}}$ is the number of times in which the solver achieved final energy lower than val\_ref, and $n_{\textrm{tot}}$ is the number of runs.
\item Method for computing the \textit{Time-To-Solution} (TTS), if the proper flag for saving the solver execution time was set. It is a figure of merit commonly employed for comparing quantum and quantum-inspired approaches with classical solutions \cite{hen2015probing,mandra2016strengths,albash2018demonstration,kowalsky20223,zielewski2022method}. It is defined as the time required to find a \textit{target solution}, which is the optimal one or a sub-optimal with final energy lower than a certain value with a percentage of confidence $p_{\textrm{conf}}$ and can be computed as: 
\begin{equation}
\vspace{-2pt}
     \textrm{TTS}  = t_{f} \frac{\log{(1-p_{\textrm{conf}})}}{\log{(1- p_{\textrm{range}}(t_{f}))}} \, ,
     \vspace{-2pt}
\end{equation}
where $t_{f}$ is the algorithm execution time, $p_{\textrm{range}}(t_{f})$ is the probability of finding energy lower than a certain value, executing the algorithm for a time $t_{f}$.
\item Method for saving all the information related to the solution in a \textit{JavaScript Object Notation} (JSON) file, allowing the \mbox{post-processing} or the exploitation of the solution at a later time. 
\end{itemize}
\examplethree{In this case, the expected optimal solution in binary variables is:
\begin{align*}
    \{b_0: 0, b_4: 1, b_1: 0, b_5: 0, b_6: 1, b_7: 1, b_8: 0, b_2: 0,\\ b_3: 0, \_a_0: 0, \_a_1: 0, \_a_2: 0, \_a_3: 0, \_a_4: 0.0\} \,
\end{align*}
while written with the original variables is 
\begin{align*}
    \{a: 0.0, b: 3.0, c: -1.0\} \,
\end{align*} The best energy is equal to $-2.0$.\\
It is possible to notice that the solution satisfies both constraints:
\begin{align*}
   b_1 + b_2 + b_4 == 1 \rightarrow 1 = 1   \, ,
\end{align*}
and 
\begin{align*}
b + c \geq 2  \rightarrow 2.0 \geq   2.0 \, .
\end{align*}
}
\vspace{-3pt}
\subsection{Penalty Weight Update}\label{sec:lambdaupdate}
\vspace{-3pt}
Finally, the framework provides further instruments, i.e., the possibility of automatically performing the constraints satisfaction check on the best solution and, if the solution is not valid, executing the algorithm again, \textit{adjusting} the value of $\lambda$ until a valid solution is found or the maximum number of trials $t$, selected by the user, is exceeded. This could be necessary, even if the tool provides methods for $\lambda$ estimation, since they are not exact methods, and consequently, their effectiveness could be problem-dependent, thus requiring adjustment.  

The framework supports three methods, presented in \cite{garcia2022exact}, for this purpose: 
\begin{itemize}
    \item \textit{sequential scaling} (sequential), which multiplies the old $\lambda$ for a factor equal to 10, increasing it by one order of magnitude;
    \item \textit{scaled-sequential scaling} (scaled), which increases the $\lambda$ with the following equation:
    \begin{equation*}
        \lambda_{\textrm{new}} = \textrm{round}(\lambda \lambda_\textrm{max}^{\frac{1}{t-1}}) \, ,
    \end{equation*}
    where$\lambda_\textrm{max}$ is the upper bound for the penalty weight;
    \item \textit{binary-search scaling} (binary search), which increases the $\lambda$ with the following equation:
    \begin{equation*}
        \lambda_{\textrm{new}} = \textrm{round}(\sqrt{\lambda \lambda_\textrm{max}}) \, .
    \end{equation*}
\end{itemize}
\vspace{8pt}
\examplethree{Finally, the problem specification of the considered example can be provided to the framework with the following code. }
\begin{lstlisting}[ language=Python]
var = Variables()
a = var.add_binary_variable("a")
b = var.add_discrete_variable("b", [-1, 1, 3])
c = var.add_continuous_variable("c", -2, 2, 0.25)
obj_func = ObjectiveFunction()
obj_func.add_objective_function( a + b * c + c**2)
cst = Constraints()
cst.add_constraint("b + c >= 2", variable_precision=True)
prb = Problem()
prb.create_problem(var, cst, obj_func)
\end{lstlisting}
\textit{To solve it with, for example, the Dwave QA, it is sufficient to write the command in the following. The framework automatically performs problem translation and the solver execution, providing the solution object previously described as output. }
\begin{lstlisting}[ language=Python]
solution = Solver().solve_dwave_quantum_annealer(prb,token=token)
\end{lstlisting}
\begin{figure*}[t]
\begin{subfigure}[t]{0.66\columnwidth}
	    \centering

\includegraphics[width=\textwidth]{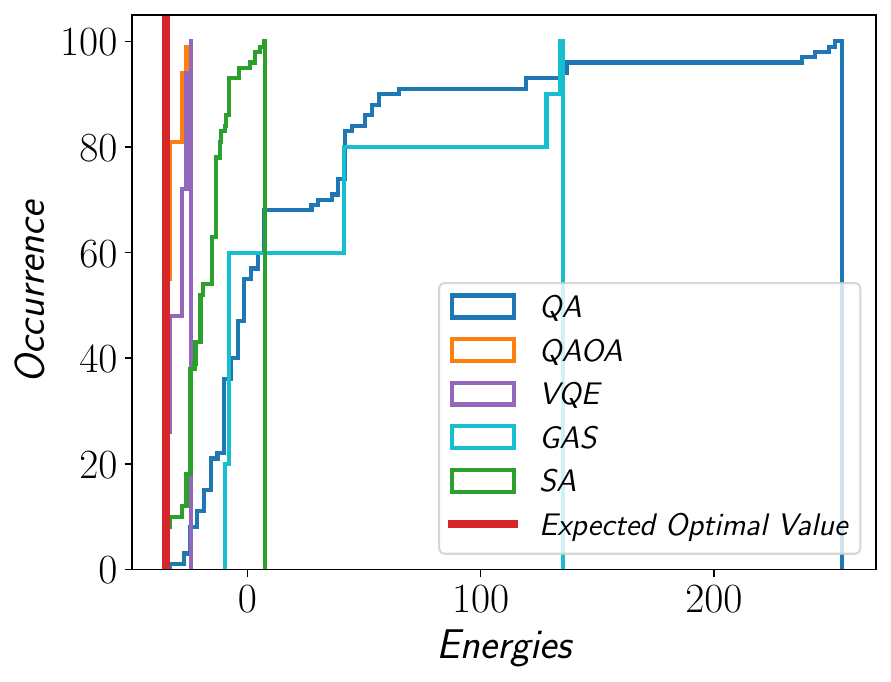}
	    \caption{ Cumulative distribution of the f3\_l-d\_kp\_4\_20 knapsack problem of the \href{http://artemisa.unicauca.edu.co/~johnyortega/instances_01_KP/}{0/1 set}, considering VLM method for $\lambda$ estimation }
	    \label{fig:KnapsackCumulative}
	\end{subfigure}
 \quad
	\begin{subfigure}[t]{0.66\columnwidth}
	    \centering
	    \includegraphics[width=\textwidth]{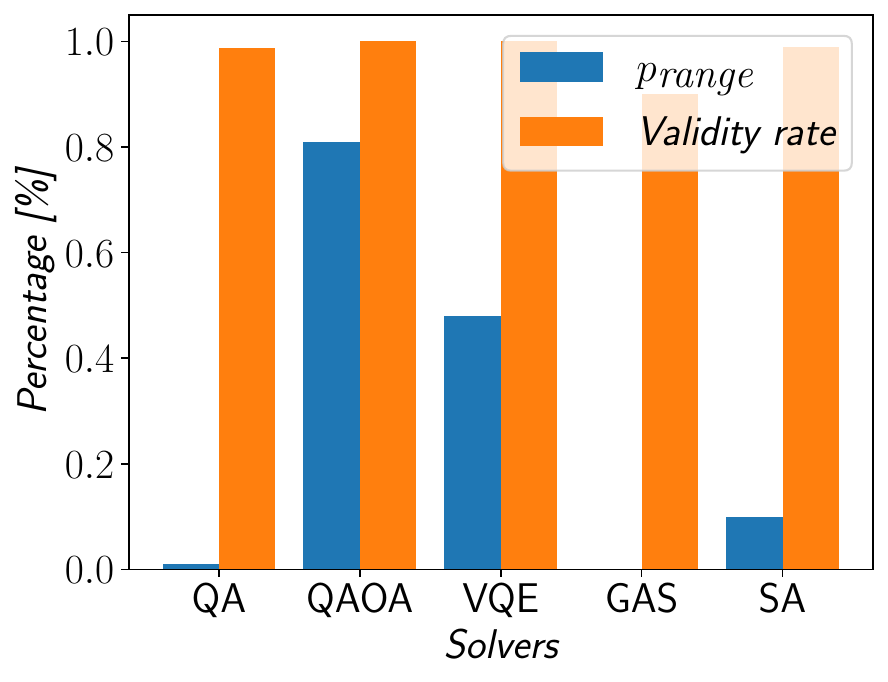}
	    \caption{Rate of valid solution and $p_\textrm{range}$ with reference value -30 (Equation \ref{eq:p_range}) for the f3\_l-d\_kp\_4\_20 knapsack problem of the \href{http://artemisa.unicauca.edu.co/~johnyortega/instances_01_KP/}{0/1 set}, considering an expected optimal energy of \mbox{-35} }
	    \label{fig:KnapsackPercentage}
	\end{subfigure}
 \quad
        \begin{subfigure}[t]{0.66\columnwidth}
	    \centering
	    \includegraphics[width=\textwidth]{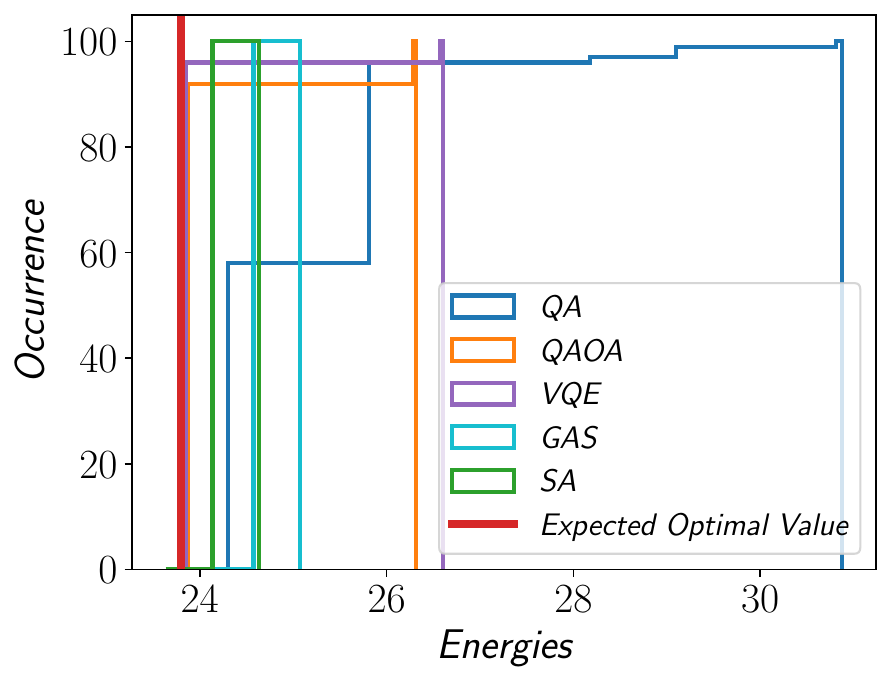}
	    \caption{Cumulative distribution of linear regression performed considering the \href{https://scikit-learn.org/stable/auto_examples/datasets/plot_iris_dataset.html}{Iris dataset}, considering two features and $w$ range $[-0.25, 0.25]$, using a precision of 0.25}
	    \label{fig:LinearRegression}
	\end{subfigure}
	\caption{Analysis of the results obtained by solving with each supported optimizer the use case problems. \vspace{-15pt} }
	\label{fig:UseCase}
\end{figure*}

\vspace{-15pt}
\section{Use cases} \label{sec:usecases} 
\vspace{-3pt}
This section summarizes the results obtained from the use cases, evaluating the proposed framework from the user's perspective in terms of effectiveness and flexibility. To this end, the case studies already covered above have been considered: the knapsack problem (chosen to showcase the handling of inequality constraints) and a linear regression task (highlighting the utilization of real variables). The problems can be solved in both instances without requiring expertise in quantum-compliant formulation and with minimal familiarity with quantum solvers.

All details of these case studies, as well as an open-source implementation of the proposed framework, are publicly available on GitHub (\url{https://github.com/cda-tum/mqt-qao}).
\vspace{-3pt}
\subsection{Knapsack}
\vspace{-3pt}
As previously mentioned, the \textit{knapsack} problem \cite{salkin1975knapsack} aims to define for a set of $N_{\textrm{obj}}$ objects, characterized by weights $w_{\textrm{arr}_i}$ and of preference scores $p_{\textrm{arr}_i}$, the best subset to put into a bag, ensuring that the total weight does not exceed the maximum $W_\textrm{max}$. The problem is conventionally formulated by associating a variable for each object ($\textrm{obj}_i$), assuming 1 if the object is in the subset and 0 otherwise, and optimizing the objective function
    \vspace{-10pt}
\begin{equation*}
    \textrm{maximize} \quad f(\textrm{obj}) = \sum_i^{N_{\textrm{obj}}} p_{\textrm{arr}_i} \textrm{obj}_i = \textrm{obj}^{T} p_\textrm{arr} \, ,
    \vspace{-3pt}
\end{equation*}
subject to
\vspace{-10pt}
\begin{equation}
    \sum_i^{N_{\textrm{obj}}} w_{\textrm{arr}_i} \textrm{obj}_i =  \textrm{obj}^{T} w_\textrm{arr} \leq W_\textrm{max} \, .
    \vspace{-3pt}
\end{equation}

Traditionally, writing the problem according to QUBO formalism usually requires following the methodology outlined in \cite{quintero2021characterizing}, consisting of the steps reviewed in Section \ref{sec:QuantumOptimization}. However, leveraging the proposed framework substantially streamlines the process. In fact, only the following code is required to provide the problem specification:
\begin{lstlisting}[language=Python]
var = Variables()
obj = vars.add_binary_variables_array("obj", [N_obj])
obj_func = ObjectiveFunction()
obj_func.add_objective_function( np.dot(np.transpose(obj), p_arr), minimization=False)
cst = Constraints()
cst.add_constraint(str(np.dot(np.transpose(obj), w_arr)) + " <= " + format(W_max))
prb = Problem()
prb.create_problem(var, cst, obj_func)
\end{lstlisting}
Afterwards, the problem can be solved with the QAOA optimizer (as one possible representative solver) by simply executing the following command:
\begin{lstlisting}[language=Python]
solution = Solver().solve_qaoa_qubo(prb)
\end{lstlisting}

For providing examples of the results analysis that can be conducted with the framework, Figures~\ref{fig:KnapsackCumulative} and~\ref{fig:KnapsackPercentage} were generated by considering the f3\_l-d\_kp\_4\_20 knapsack problem of the \href{http://artemisa.unicauca.edu.co/~johnyortega/instances_01_KP/}{0/1 set}, which comprises the weights and preference scores for different sets of objects. In particular,  Figure~\ref{fig:KnapsackCumulative} shows the cumulative distribution derived from executing each supported solver one hundred times on the problem, while Figure~\ref{fig:KnapsackPercentage} illustrates the rate of valid solutions, i.e., those satisfying the constraint, and the $p_\textrm{range}$ values computed by the framework using Equation~\ref{eq:p_range}, with a reference value of -30, proximate to the expected optimal value of -35. 


\vspace{-3pt}
\subsection{Linear Regression}
\vspace{-3pt}
\textit{Linear regression} \cite{date2021qubo} is a statistical technique for modeling the relation between variables and others exploitable in a wide range of applications, including machine learning. As discussed above, it consists of identifying the straight line that best fits some available data. The model to optimize can be expressed as
\begin{equation*}
    \textrm{minimize} \quad E(w) = \Vert X w + Y \Vert \, ,
\end{equation*}
where $X \in \mathbb{R}^{N\times(d+1)}$ is the augmented regression data matrix, $N$ the number of data points in the training set, $d$ the number of features, $Y \in \mathbb{R}^{N}$ is the regression label of the training data, $w \in \mathbb{R}^{d+1}$ the regression weights and $E(w)$ the Euclidian error function. This can be rewritten in a linear form as
\begin{equation*}
    \textrm{minimize} \quad E(w) = w^{T}X^{T}X^{T}Xw - 2w^TX^TY+Y^TY  \, ,
\end{equation*}
allowing the exploitation of the framework for describing the problem---establishing the operative range (minv, maxv, precision) of the regression weights---in a practical way, as shown in the following, instead of applying the complex procedure presented in \cite{date2021qubo}, consisting of the steps discussed in Section \ref{sec:QuantumOptimization}. 

More precisely, the problem specification can be provided to the framework with the following instructions:
\begin{lstlisting}[language=Python]
var = Variables()
w = var.add_continuous_variables_array("w", [d + 1], minv, maxv, precision)
obj_func = ObjectiveFunction()
obj_func.add_objective_function(np.dot(np.dot(np.dot(np.transpose(w), np.transpose(X_training)), X_training), w) - 2 * np.dot(np.dot(np.transpose(w), np.transpose(X_training)), Y_training)+ np.dot(np.transpose(Y_training), Y_training),)
cst = Constraints()
prb = Problem()
prb.create_problem(var, cst, obj_func)
\end{lstlisting}
Afterwards, the VQE optimizer (as one possible representative solver) can solve it by launching the following command:
\begin{lstlisting}[language=Python]
solution =  Solver().solve_vqe_qubo(prb)
\end{lstlisting}

For showing the framework tools for results analysis,  Figure \ref{fig:LinearRegression} shows the cumulative distribution obtained by executing one hundred times each supported solver considering two features of the \href{https://scikit-learn.org/stable/auto_examples/datasets/plot_iris_dataset.html}{Iris dataset}  and $w$ range $[-0.25, 0.25]$ with precision 0.25.

\subsection{Discussion} \label{sec:discussion}
The two real-world use cases discussed demonstrate the framework's comprehensiveness, adeptly managing constraints and real variables. Observing the code implementations, it becomes evident that the user interface is intuitive and closely resembles conventional optimization packages, enabling users to express their problems in a familiar and conventional way based on variables, constraints, etc.  A single black box function completely hides the complexities of translating problems into QUBO formalism and harnessing quantum solvers from non-expert users. This function produces a solution object common to all solvers as output.

For users possessing a certain level of expertise, the framework offers the flexibility to manually specify various aspects, such as the encoding method or the mechanism for~$\lambda$ estimation, as well as the solver parameters, rather than relying on default settings. Furthermore, as shown in Figure~\ref{fig:UseCase}, the framework provides valuable tools for analyzing the obtained solution and facilitating comparisons between different solvers.
\vspace{-5pt}
\section{Conclusions}\label{sec:conclusions}
\vspace{-3pt}
This work introduced a framework, publicly available on GitHub (\url{https://github.com/cda-tum/mqt-qao}) as part of the Munich Quantum Toolkit (MQT), designed to empower non-experts in quantum computing and QUBO formulation to explore the potential of quantum solutions for optimization problems. To this end, we first reviewed the procedure required for solving an optimization problem with a quantum solver. Then, we demonstrated how the framework automates this process, providing users with an interface as similar as possible to one of the commonly used packages for conventional optimization. This automation conceals the complexity of formulating problems according to the QUBO formalism and managing quantum solvers. The effectiveness and flexibility of the framework have been shown in two different case studies: the knapsack problem and linear regression modeling. In both cases, it could be seen that, after the declaration of the problem specifications, a solution can be obtained by executing a simple black-box command.

Even though the considered use case has shown the potential and the resulting benefit of the framework for the users, opportunities for enhancement and expansion remain. First, partial or total automation of quantum solver parameter adaptation tailored to specific problems could enhance efficiency. Moreover, the framework can be exploited for performing comparisons among quantum solvers and developing a prediction mechanism to identify the most suitable solver for a given problem, automating the solver selection. Finally, recognizing that the main limitation of the tool is the requirement of providing input cost functions in a polynomial form---excluding non-linearities such as exponential or trigonometric functions---a linearization step can be added, thus expanding its application domain.

In conclusion, the proposed framework lays the foundation for automating quantum computing solutions and broadening access to quantum optimization for a wider range of users.
\vspace{-5pt}
\section*{acknowledgements}
\vspace{-5pt}
N.Q. and R.W. acknowledge funding from the European Research Council (ERC) under the European Union’s Horizon 2020 research and innovation program (grant agreement No. 101001318), the Munich Quantum Valley, which is supported by the Bavarian state government with funds from the Hightech Agenda Bayern Plus, and the BMWK on the basis of a decision by the German Bundestag through project QuaST, as well as the BMK, BMDW, the State of Upper Austria in the frame of the COMET program, and the QuantumReady project within Quantum Austria (managed by the FFG).

\vspace{-5pt}
\bibliographystyle{ieeetr}
\bibliography{sn-bibliography}

\end{document}